\documentstyle[aps,prl,preprint]{revtex}

\begin{document}
\title{A Unified Theoretical Description of the Thermodynamical Properties of Spin
Crossover with Magnetic Interactions}
\author{Kamel Boukheddaden$^{1}$, Masamichi Nishino$^{2,*}$, Seiji Miyashita$^{3}$,
and Fran\c cois Varret$^{1}$}
\address{$^{1}${\it Laboratoire de Magn\'{e}tisme et d'Optique, CNRS-Universit\'{e}}\\
de Versailles/St. Quentin en Yvelines 45 Avenue des Etats Unis, F78035\\
Versailles Cedex, France \\
$^{2}${\it Computational Materials Science Center, National Institute for}\\
Materials Science,\\
Tsukuba, Ibaraki 305-0047, Japan \\
$^{3}${\it Department of Physics, Graduate School of Science,}\\
The University of Tokyo, Bunkyo-Ku, Tokyo, Japan.}
\date{\today}
\maketitle

\begin{abstract}
After the discovery of the phenomena of light-induced excited spin state
trapping (LIESST), the functional properties of metal complexes have been
studied intensively. Among them, cooperative phenomena involving low
spin-high spin (spin-crossover) transition and magnetic ordering have
attracted interests, and it has become necessary to formulate a unified
description of both phenomena. In this work, we propose a model in which
they can be treated simultaneously by extending the Wajnflasz-Pick model
including a magnetic interaction. We found that this new model is equivalent
to Blume-Emery-Griffiths (BEG) Hamiltonian with degenerate levels. This
model provides a unified description of the thermodynamic properties
associated with various types of systems, such as spin-crossover (SC) solids
and Prussian blue analogues (PBA). Here, the high spin fraction and the
magnetization are the order parameters describing the cooperative phenomena
of the model. We present several typical temperature dependences of the
order parameters and we determine the phase diagram of the system using the
mean-field theory and Monte Carlo simulations. We found that the magnetic
interaction drives the SC transition leading to re-entrant magnetic and
first-order SC transitions.
\end{abstract}


\section{Introduction}

Recently the functionality of metal complexes attracts much interest. In
particular, the cooperative phenomena where both spin-crossover (SC) and
magnetic ordering are involved have been studied intensively. Molecular
switching of inorganic solids is a typical issue of the vibronic lability of
molecular units, originally introduced in chemistry by J.A.\ Ammeter \cite
{Ammeter,Kahn}.

The concepts of molecular bistability were given by O.\ Kahn \cite{Kahn},
and recent reviews on this subject can be found in the references\cite
{Varret,Sorai,Gutlich,varret2}. In addition, due to the characteristic of
reversible control of the magnetic properties by light, temperature,
pressure, and magnetic field, SC and PBA compounds are considered to be
promising candidates for information storage.\cite{Verdaguer,Sato1,Freysz}. 
\[
\begin{array}{llll}
\text{system} & \text{low-temperature sate} & \text{High temperature State}
& \text{ref.} \\ 
\text{SC system Fe}^{\text{II}}\text{ (3d}^{\text{6}}\text{)} & \text{Fe}_{%
\text{LS}}^{\text{II}}\text{ (S=0)}\;[\text{t}_{\text{2g}}^{6}] & \text{Fe}_{%
\text{HS}}^{\text{II}}\text{ (S=2)}\;[\text{t}_{\text{2g}}^{4}\text{e}_{%
\text{2g}}^{2}] & \text{\cite{tof,Koenig,gut2}} \\ 
\text{PBA Fe,Co } & \text{Fe}_{\text{LS}}^{\text{II}}\text{ (S=0) Co}_{\text{%
LS}}^{\text{III}}\text{(S=0)} & \text{Fe}_{\text{LS}}^{\text{III}}\text{%
(S=1/2) Co}_{\text{HS}}^{\text{II}}\text{(S=3/2)} & \text{\cite
{Sato,Bleuzen,Goujon}}
\end{array}
\]

In SC, the electronic structure changes between the high spin state (HS) and
and the low spin state (LS) in an atom which causes changes of magnetic
property and also volume of the molecule. On the other hand, in PBA the
charge transfer between Fe and Co causes a LS-HS (spin-crossover) transition
of Co, and the magnetic moments of the Co and Fe couple
antiferromagnetically in the high spin state (HS), which results in local
ferrimagnetic moment. Although the LS-HS transition of PBA is different from
simple SC transitions, we can regard PBA as a SC system in the broad sense.
A new aspect of PBA is the magnetic ordering at low temperatures, where the
magnetic interaction between spins is important.

As another type of phase transition due to the charge transfer, other new
inorganic systems, showing magnetic changes after the spin transition, have
been reported in literatures, such as mixed-valence iron complex (nC$_{3}$H$%
_{7}$)$_{4}$N[Fe$^{\text{II}}$Fe$^{\text{III}}$(dto)$_{3}$](dto=C$_{2}$O$%
_{2} $S$_{2}$) \cite{Kojima,Miya-Kojima}, where the charge transfer causes a
change of degeneracy, and a phase transition occurs between a high
temperature and low temperature states. In this system, a magnetic
transition is also observed at a low temperature. There, the spin-crossover
of each atom does not occur. However, the entropy effect plays an important
role in the structure change associated with the electronic configuration.
It is an interesting problem to study the mechanism of combined phenomena of
the structure change and magnetic transition.

In SC transition, a cooperative distortion of the lattice occurs, which is
associated with a large  change in the average metal-ligand distance. Indeed
it increases about $10\%$ at the molecular level in the conversion from the
LS to HS state which causes a volume expansion of the molecules. Thus, the
SC phenomenon at the microscopic scale is due to the coupling between the
electronic and the vibrational structures \cite{Kambara} of the molecules.
That is the intra-molecular vibronic coupling. The coupling between the
lattice distortion and the change of the electronic states of the molecules
induces the cooperativity of SC solids. This is the microscopic origin of
the elastic interaction which is responsible for the SC transition. From the
experimental point of view, optical and magnetic properties change during
the SC transition, so that the switching properties can be followed by
magnetic or optical (absorption, reflectivity) techniques.

For the analysis of these structural transition, it is largely admitted that
the bistable properties at the molecular level are adequately described
through a molecular configurational diagram, i.e. a plot of adiabatic
energies versus distortion coordinate of the molecular system. In Fig.\ref
{fig-energylevel}, we show a unified three-states configurational diagram
suited to SC and PBA in the case of the low spin (LS) ground state. It is
noteworthy that the effect of environment in molecular solids affects the
configurational diagram. For example, an external pressure mainly increases
the $E($HS$)-E($LS$)$ energy gap, 
and thus reduces energy barrier between HS and LS states in SC solids.
Consequently, it raises the transition temperature ($T_{{\rm eq}}$) and it
decreases the lifetime of the metastable HS state \cite
{Jeftic,Boillot,Jeftic2}. In the case of PBA, the same effect of pressure is
also expected on the spin transition 
of cobalt \cite{AncaSava}, which would cause variation of the magnetic
properties under high pressures. In the mixed-valence iron complex (nC$_{3}$H%
$_{7}$)$_{4}$N[Fe$^{\text{II}}$Fe$^{\text{III}}$(dto)$_{3}$](dto=C$_{2}$O$%
_{2}$S$_{2}$
), the pressure effect has been reported \cite{Kojima2}.

\begin{figure}[h]
\caption{A unified three-states configurational diagram. The non-degenerate
ground state is of low spin and the excited degenerate states are of high
spin.}
\label{fig-energylevel}
\end{figure}

To describe SC systems, a two-states Ising-like model proposed by Wajnflasz
and Pick (WP) \cite{Pick} is currently used. In this phenomenological model,
the HS and LS microscopic states of the molecule are associated,
respectively, with the eigenvalues $+1$\ and $-1$\ of the fictitious spin
operator $s$ with their respective degeneracies $g_{+}$\ and $g_{-}$. Here,
we denote by $g=\frac{g_{+}}{g_{-}}>1$\ the degeneracy ratio, by $\Delta >0$%
\ the ligand field energy (such that the LS state is the ground state at $0$%
K), and by $J^{\prime }$\ the interaction between the molecules. $J^{\prime }
$\ is assumed to have an independent value on the spin states. Taking into
account the isomorphism \cite{Doniach} between the Ising model with
degenerate levels and an Ising model under a temperature dependent ''field'' 
\cite{kbo}, the WP Hamiltonian is written as 
\begin{equation}
H=-J^{\prime }\sum_{i\neq j}s_{i}s_{j}+\left( \Delta -\frac{k_{\text{B}}T}{2}%
\ln g\right) \sum_{i}s_{i},  \label{WP}
\end{equation}
where $\sum\limits_{i\neq j}$ is the sum over the interacting neighbors and 
$s$ takes 1 (HS) or -1 (LS). $%
J^{\prime }>0$ is the interaction parameter between the SC units.

As we mentioned before, the interaction in SC solids originates from elastic
strength \cite{Spiering1}.\ However, in the WP model it is expressed through
the phenomenological term $-J^{\prime }\sum\limits_{i\neq j}s_{i}s_{j}$.
Here $J^{\prime }$ is not an exchange magnetic coupling. This term expresses
the situation that an attractive force (i.e. a ferroelastic interaction)
acts between neighboring sites in the same state (LS-LS or HS-HS) and a
repulsive force acts between those in different states (LS-HS), which leads
to bistability of the LS and HS states. A more realistic treatment which
includes a spin-phonon interaction leads to obtain a similar Hamiltonian as (%
\ref{WP}) with effective parameters $J^{\prime }$ and $\Delta $\ which
depend on the distortions of the molecules at sites $i$\ and $j$\ \cite
{Jamil1}. Thus, it is important to consider here the parameters $J^{\prime }$%
\ and $\Delta $\ as effective parameters which have elastic origin, even if
they are taken as constants in the model in order to make the present
treatment easy.

The degeneracy ratio ($g$) between the HS and LS states is related to the
molar entropy change ($\Delta S$) due to the total spin-conversion which is
given by $\Delta S=Nk_{\text{B}}\ln g$ , where $k_{\text{B}}$ is the
Boltzmann constant and $N$\ the Avogadro's number. Because of both
electronic, vibrational (intramolecular) and phonon factors, the
experimental entropy changes by $\Delta S=30-70$JK$^{-1}$mol$^{-1}$\cite
{Sorai}. This corresponds to the following degeneracy values $g=36-4500$,
which indicates that $g\gg 1.$ Generally, $g$\ depends on temperature
through the temperature dependence of the phonon density of the lattice \cite
{Jamil}. Here, we simply take $g$\ as a constant.

The ratio of molecules in the high spin state, called HS fraction, is
denoted by $n_{\text{HS}},$ which is expressed as a function of the
``fictitious magnetization'' $<s>$ such as $n_{\text{HS}}=\left(
<s>+1\right) /2.$ Despite of the drastic simplification of the realistic
vibronic levels scheme by the two-level system, this model permits to
describe with amazing success most of the quasi-static properties of the SC
solids. For example, it explained consistently the change from a smooth
transition to a first order transition\cite{Pick,kbo,nishino3} of spin
conversion between the LS and HS states as a function of the parameters of
the model. It also well explained the two-step spin crossover behavior \cite
{Az1,koppen,nishino4}.

In order to study the magnetic properties in PBA solids, etc., we need to go
beyond the 2-states model. In previous studies \cite{nishino1,nishino2},
focusing on the condition of the reversible switching between nonmagnetic LS
state and the magnetic metastable ordered HS state which was observed in PBA%
\cite{Sato}, the following three state model called Blume-Capel model \cite
{BC1,BC2} was investigated: 
\begin{equation}
H=-J\sum\limits_{i\neq j}S_{i}S_{j}+\Delta \sum\limits_{i}S_{i}^{2}.
\label{BCapel}
\end{equation}
In this model, the spin operator $S_{i}$ has three states: $-1,0,+1$, where
0 ($\pm 1$) is associated with the LS nonmagnetic (HS magnetic) state. Here, 
$J$ is really an exchange magnetic coupling and $\Delta $ plays a role of a
ligand field. Reversible switching was demonstrated by a suitable change of
parameters of the model assuming photon's effect is renormalized into the
parameters. In addition, from a dynamical viewpoint, it has been pointed out
that existence of multi-time scale dynamics is essential for the switching
in both directions (magnetic $\leftrightarrow $ diamagnetic). That is, the
difference between the time scales of the spin flip process and of
structural change (LS$\leftrightarrow $HS) process is important. This
concept is very important to investigate photomagnetism of PBA\cite{next}.

To reproduce the temperature-induced SC properties of PBA, the information
of the difference in degeneracies between the high temperature and the low
temperature phases and the nonmagnetic interaction between molecules (which
is originated from the lattice) have to be taken into account. This is the
motivation of this paper. Here, we propose a model taking into account both
mechanisms of the temperature-induced SC transition and of the magnetic
transition. Then, we extend the 3-states model \cite{nishino1,nishino2}
(given by the Hamiltonian (\ref{BCapel})) to include the mechanism in the
Wajnflasz and Pick model (Hamiltonian (\ref{WP})). As we will see in the
next section, this  present model can be regarded as an extended
Blume-Emery-Griffiths (BEG) Hamiltonian \cite{BEG}.

It is worth noting that for the elastic interaction between molecules,
alternative scenarios to describe the SC phenomenon have been proposed
recently. Some of them are based on pseudo-spins coupled by springs \cite
{Jamil1}. The originality of this approach is that the elastic constant
between two neighboring atoms depends on their electronic states. Indeed,
due to the expansion of the lattice in the HS state, the elastic force
constants are weaker in the HS state than in the LS state. This approach
leads to a Hamiltonian which is equivalent to an Ising-like model in which
the interaction parameter depends on the local distortions of the molecules.
An extension of this scenario including the intramolecular potential energy
of the molecules has been published recently by one of the authors \cite
{PTP_Kamel}. However, due to the complicated structure of the Hamiltonians,
these models have been investigated only in the one dimensional case,
but we expect new development of this kind of descriptions in near future.

This paper is organized as follows: in section II we introduce the model,
section III is devoted to study the equilibrium properties obtained 
by a mean-field theory. In section IV, results of Monte Carlo simulations
are presented. In section V, we discuss the obtained results in 
comparison with available experiments in the literature and we conclude.

\section{The Hamiltonian}

Here we adopt a model based on a three states Ising-like Hamiltonian of $%
S=0,+1,-1.$ The state $S=0$ denotes the non-magnetic state, which
corresponds 
to the electronic configuration Fe$_{\text{LS}}^{\text{II}}$Co$_{\text{LS}}^{%
\text{III}}$ in the case of PBA, and simply Fe$_{\text{LS}}^{\text{II}}$ for
SC systems. The states $S=\pm 1$ are associated with magnetic states. The
magnetic state corresponds to the ferrimagnetic state Fe$_{\text{LS}}^{\text{%
III}}$Co$_{\text{HS}}^{\text{II}}$ for PBA compounds, and Fe$_{\text{HS}}^{%
\text{II}}$ in the case of SC solids. Due to large expansion of the
coordination sphere of the molecule during the spin transition, the
vibrational structure \cite{Sorai} and the electronic properties between the
HS and LS states are strongly affected. Consequently, the degeneracies of
the non-magnetic state and the magnetic state are different. Let us denote
by $u$ and $r$ the degeneracies of the states $S=0$ and $S=\pm 1$,
respectively, and by $g$ the ratio $\frac{r}{u}$. Here we assume that the
degeneracies of the states with $S=\pm 1$ are the same and $g$\ is
temperature independent.

Because the high and low spin states are described by $S=\pm 1$ and $S=0$,
respectively. The Wajnflasz and Pick model (\ref{WP}) in the present three
states approach, in which the degeneracies are included as entropic terms 
\cite{Pick,kbo,Doniach,PRB}, is given by 
\begin{equation}
H_{{\rm WP}}=-J^{\prime }\sum_{i\neq j}(2S_{i}^{2}-1)(2S_{j}^{2}-1)+\left(
\Delta -\frac{k_{\text{B}}T}{2}\ln g\right) \sum_{i}(2S_{i}^{2}-1).
\label{HamWP}
\end{equation}
Adding the magnetic interaction $-J\sum_{i\neq j}S_{i}S_{j}$ and the Zeemann
energy $-h\sum_{i}S_{i}$, we obtain the general Hamiltonian: 
\begin{equation}
H=-J\sum_{i\neq j}S_{i}S_{j}-h\sum_{i}S_{i}-K\sum_{i\neq
j}S_{i}^{2}S_{j}^{2}+\left( D-k_{\text{B}}T\ln g\right) \sum_{i}S_{i}^{2}+%
\text{const{\bf .}}  \label{Ham}
\end{equation}
where $K=4J^{\prime }$ ${\bf (>}0{\bf )}$ and $D=2\Delta +4zJ^{\prime }$ $%
{\bf >}0$ ($z$ : the coordination number). Here only the nearest neighbor
interactions are considered for the pairing $i\neq j$. The magnetic
interaction between the 
spins is only relevant when both spins $S_{i}$ and $S_{j}$ are in the
magnetic state, i.e. $S=\pm 1$. If we do not consider the degeneracies,
i.e., $u=r=1$\ and $g=1$, this Hamiltonian corresponds to the well known BEG
model.\cite{BEG} Here, we adopt the form of Eq. (\ref{Ham}) to refer to past
works on BEG model. Indeed, BEG Hamiltonian was extensively studied to
describe the thermodynamical properties of the tri-critical point in He$^{%
{\bf 3}}$-He$^{4}$\ mixtures \cite{BEG}. Later, Sivardi\`{e}re and
Lazerowitch \cite{Sivardiere} used an extended version to lattice gas model
of BEG Hamiltonian to study the condensation and phase separation in binary
fluids. It is also interesting to mention that even recently, BEG
Hamiltonian was also applied \cite{Luty} to describe the ferroelectric order
thermally induced in donor-acceptor materials like TTF-TCNQ \cite{Koshihara}%
. This kind of materials are also intensively studied for their
photo-induced properties \cite{Koshihara2}.

\section{Equilibrium properties}

In experiments, so far the spin crossover and magnetic ordering in PBA take
place in the range of temperature $10$ $-300$ K. 
As typical values of the parameter of the model, extracted from the 
experiments \cite{Goujon,Varret3}, we can refer to the exchange interaction $%
zJ\simeq 20$K , the ligand-field energy gap $D\approx 500$ K, the elastic
interaction $zK$ is about $100-200$K and the degeneracy ratio{\bf \ }$%
g\simeq 100-200.$ Here $z$ is the number of the nearest neighbor pairs. The
present model describes the systems of SC and PBA solids as special cases:

(i) for SC systems, the magnetic interaction between the magnetic HS units
is negligible because of the large distance between them ($\approx 10$ $\AA $%
) \cite{Galina}.

(ii) In the case of PBA, all the terms of the model should be taken into 
account, because they are effectively present. However, as far as we concern
the equilibrium 
thermodynamics, only the electron-transfer contributes to the thermal
bistability between the states Fe$_{\text{LS}}^{\text{II}}$Co$_{\text{LS}}^{%
\text{III}}$ and Fe$_{\text{LS}}^{\text{III}}$Co$_{\text{HS}}^{\text{II}}$.
Indeed around the spin crossover transition temperature, experimentally
around $200-300$K, there is no magnetic ordering. 
This fact suggests that the magnetic coupling $zJ,$ which is about $20$K, is
very small in comparison to the ligand-field energy gap $D\approx 500$ K.
However, in the near future, because of an amazing developments of synthesis
of materials in this subject, we may expect new systems in which the
magnetic and SC properties could compete. Thus, we study a general structure
of the phase diagram caused by $K$ and $J$ in the present paper.

\subsection{The SC case}

As we mentioned above, in the case of SC, the magnetic interaction is
irrelevant and the Hamiltonian (\ref{Ham}) without the magnetic field and
magnetic interaction can be written in the following form 
\begin{equation}
H_{eq}=-K\sum_{i\neq j}S_{i}^{2}S_{j}^{2}+(D-k_{\text{B}}T\ln
g)\sum_{i}S_{i}^{2}.  \label{HamEq}
\end{equation}
It is straightforward to see that a flipping of any spin $\left(
S_{i}\rightarrow -S_{i}\right) $ does not change the Hamiltonian. Therefore,
as expected, the magnetization $m=<S>$ should be zero at any temperature.
Introducing a new variable $\sigma _{i}=2S_{i}^{2}-1$, the Hamiltonian \ref
{HamEq} reduces to WP model\cite{Pick}, which leads to a first order phase
transition as it is well known from the literature\cite
{kbo,nishino3,LSV,1Dkbo}: 
\begin{equation}
H_{\sigma }=-\frac{K}{4}\sum_{i\neq j}\sigma _{i}\sigma _{j}+\frac{D-k_{%
\text{B}}T\ln 2g-zK/2}{2}\sum_{i}\sigma _{i}+\text{const.}  \label{Hsigma}
\end{equation}
In a mean-field approximation, the order parameter of the system is $%
q=<S^{2}>$. It corresponds exactly to the HS fraction of molecules in the
system. The self-consistent equation accounting for the thermally driven
first-order transition between the macroscopic LS ($q\simeq 0$) and HS ($%
q\simeq 1$) states, is given by 
\begin{equation}
k_{\text{B}}T=\frac{D-zKq}{\ln \left( \frac{2g(1-q)}{q}\right) }.
\label{selfEq2}
\end{equation}
So-called transition temperature $T_{{\rm eq}}$ is obtained by setting $q=1/2
$ in the equation (\ref{selfEq2}), or more simply by setting the effective 
field $D-k_{{\rm B}}T\ln 2g-zK/2$ of the Hamiltonian (\ref{Hsigma}) equal to
zero. Thus we have 
\begin{equation}
k_{\text{B}}T_{{\rm eq}}=\frac{D-zK/2}{\ln 2g}.  \label{kTeq}
\end{equation}

It is interesting to notice that occurrence of the first-order transition is
only possible when the condition $T_{{\rm eq}}<T_{{\rm I}}$, where $T_{{\rm I%
}}$ denotes the critical temperature of the pure spin $1/2$ Ising model in
the same lattice. In the mean-field 
approximation, the Curie temperature is given by $T_{{\rm I}}=zK/4k_{\text{B}%
} $.

Two types of behavior of the order parameter can be obtained in the 
mean-field approach depending on the ligand-field energy gap $D$ and the
interaction $K$:

(i) a first-order spin-crossover transition, with a hysteresis loop, when $%
zK>4D/(2+\ln 2g);$

(ii) a smooth conversion between the two states when $zK<4D/(2+\ln 2g)$.

The temperature dependence of $q$\ is summarized in Fig.\ref{fig-MFq}.

\begin{figure}[h]
\caption{Temperature dependence of the high spin fraction $q$ for values of
the energy gap $D$. From the left to the right, $D=90$K, $100$K, $120$K and $%
150$K with the fixed values of parameters $K=90$, and $g=10$. Smooth
(first-order) transitions are obtained for $D>100$K ($D<100 $K). The part of
the curves with negative slope $dq/dt$ represents unstable states.}
\label{fig-MFq}
\end{figure}


\subsection{General case.}

Here, we consider the general model of SC solids in which the magnetic
moments of the HS state interact magnetically, which may cause a magnetic 
order. Therefore, in this section we consider the Hamiltonian (\ref{Ham})
with various parameters. 
The model imposes two order parameters $m=<S>$ and $q=<S^{2}>$,
respectively, associated with the magnetization and the high spin fraction
in the system. If $K\gg J$ as the case of PBA, we can deal with a
multi-scale energy problem in the present 
frame work: high energy scale for the spin-crossover phenomena (at high
temperature), low energy scale for magnetic phenomena (at low temperature).

Now, let us study properties of the system (\ref{Ham}) by a mean-filed
theory (MFT) which gives the phase diagram of the model. The one site
Hamiltonian for this system is given by 
\begin{equation}
H_{i}=-\left( zJm+h\right) S_{i}-\left( zKq-D+k_{\text{B}}T\ln g\right)
S_{i}^{2}.  \label{HmF}
\end{equation}

The associated mean-field free energy per site is given by 
\begin{equation}
F(m,q,T)=\frac{1}{2}zJm^{2}+\frac{1}{2}zKq^{2}-k_{\text{B}}T\ln \left(
1+e^{\beta \left( zKq-D+k_{\text{B}}T\ln 2g\right) }\cosh \beta \left(
h+zJm\right) \right) .  \label{Elib}
\end{equation}
Minimizing the free energy (\ref{Elib}) with respect to the order parameters 
$m$ and $q$, we easily obtain the following coupled self-consistent
equations on $m$ and $q$, 
\begin{equation}
m=\frac{\sinh \beta \left( h+zJm\right) }{e^{-\beta \left( zKq-D+k_{\text{B}%
}T\ln 2g\right) }+\cosh \beta \left( h+zJm\right) }=q\tanh \beta \left(
h+zJm\right)   \label{mself}
\end{equation}
and 
\begin{equation}
q=\frac{\cosh \beta \left( h+zJm\right) }{e^{-\beta \left( zKq-D+k_{\text{B}%
}T\ln 2g\right) }+\cosh \beta \left( h+zJm\right) }.  \label{qself}
\end{equation}

If $J=h=0$, we reproduce naturally the case of SC previously studied.
Equations (\ref{HmF},\ref{mself},\ref{qself}) have the same structure as
that found in BEG model. However, due to the temperature dependent
ligand-field (anisotropy in BEG model) we expect new behavior for the order
parameters $m$\ and $q$\ and also a different phase diagram. Putting $g=1$
gives exactly the same self-consistent equations as that published by
Mukamel and Blume \cite{Mukamel} in their study of the BEG model to describe
the thermodynamical properties of the tricritical point in He$^{{\bf 3}}$-He$%
^{{\bf 4}}$ mixtures$.$ Our aim here is to analyze the effect of the
magnetic interaction on 
the spin-crossover phenomena ($g\gg 1$). Therefore, we study the following
two cases:

1. $J=0, h\neq 0, K\neq 0, D\neq 0;$

2. $J\neq 0,h=0,K\neq 0,D\neq 0;$

\subsubsection{$J=0,h\neq 0,K\neq 0,D\neq 0$ case}

This case corresponds to the situation of the SC system under a magnetic
field \cite{Koshi1,Azpulse}; the spins are too far to interact each other.
In such situation, the system is paramagnetic or diamagnetic, and the
magnetization has no strong effect. However, 
we study the effect of the magnetic field on the transition temperature and
on the 
hysteresis loop in detail. In the present situation, the self-consistent
equations (\ref{mself},\ref{qself}) become 
\begin{equation}
m=q\tanh \beta h  \label{m2}
\end{equation}
and 
\begin{equation}
q=\frac{\cosh \beta h}{e^{-\beta \left( zKq-D+k_{\text{B}}T\ln 2g\right)
}+\cosh \beta h}.  \label{q2}
\end{equation}

It is clear from Eqns. (\ref{m2} and \ref{q2}) that the magnetization
depends linearly on the HS fraction in the system. In the limiting case, 
where the magnetic field is very small in comparison with the transition
temperature, the magnetization is $m\simeq q\beta h.$ In the following, we
will demonstrate that the magnetic field stabilizes 
HS (magnetic) state and decreases the transition temperature $T_{{\rm eq}}$
of the system. So, we expand the equation (\ref{q2}) around the transition
temperature of the first-order transition, for which $q=1/2,$ by considering 
that $\beta h\ll 1$. After some simple calculations, we arrive at the
following expression of the field dependence of transition temperature $T_{%
\text{eq}}(h)$ 
\begin{equation}
T_{\text{eq}}(h)=T_{\text{eq}}(0)-\frac{h^{2}}{D-zK/2}.  \label{DeltaTeq}
\end{equation}
Here, $T_{\text{eq}}(0)=\frac{D-zK/2}{k_{\text{B}}\ln 2g}$ is the transition
temperature of the system under zero applied magnetic field. The shift of
the transition temperature 
\begin{equation}
\Delta T_{\text{eq}}=-\frac{h^{2}}{k_{\text{B}}T_{\text{eq}}(0)\ln 2g}=-%
\frac{h^{2}}{\Delta H(0)}  \label{DeltaTeq2}
\end{equation}
depends on the ratio between the square of the magnetic field and the
enthalpy change, $\Delta H(0)$, at the transition. 
These results are in good agreement with available experimental results and
the microscopic analysis on SC \cite{Azpulse,PRBpulse} phenomena. It follows
from 
this result that, lowering the transition temperature $T_{{\rm eq}}(0)$
enhances the magnetic field effect on SC systems.

\begin{figure}[h]
\caption{
Effect of the 'magnetic' field $h$ ($h=0$K, $h=40$K) for $K=200$K, $%
D=300$K, $g=150$) on the temperature dependence of the HS fraction $q(T)$ in
the case of the first order phase transition. 
The magnetic field stabilizes the HS phase and polarizes the magnetic system 
in which it induces a nonzero magnetization (dashed line).
}
\label{fig-MFqh}
\end{figure}

On a hysteresis loop, see Fig.\ref{fig-MFqh}, the magnetic field 
has as effect to shift the whole hysteresis to the low temperature side, by
acting differently on the ascending and descending metastable branches. 
Denoting, by $T_{h}^{+}$ ($T_{0}^{+}$) and $T_{h}^{-}$ ($T_{0}^{-}$), the 
transition temperatures under the field $h$ ($h=0$), 
at the ascending and descending branches, respectively, one can deduce
approximately from the equation (\ref{DeltaTeq}) that the associated shifts
are given 
\begin{equation}
k_{\text{B}}\Delta T^{+}=-\frac{h^{2}}{T_{0}^{+}\ln 2g}\text{ and }k_{\text{B%
}}\Delta T^{-}=-\frac{h^{2}}{T_{0}^{-}\ln 2g}.  \label{DeltaT+}
\end{equation}
Since $T_{0}^{+}>T_{0}^{-}$, it follows that $\Delta T^{-}>\Delta T^{+}$,
which indicates that the magnetic field causes a non-symmetric distortion of 
the hysteresis loop of SC solids by stabilizing the HS and destabilizing the
LS state.

To give an order of magnitude of the shift of the hysteresis loop, we
consider the case of $\text{Co}^{II}(\text{H}_{2}\text{(fsa)}_{2}\text{en})%
\text{(py)}_{2}$ $[\text{H}_{2}\text{(fsa)en}=\text{N,N'-ethylene
bis(3-carboxysalicylaldimine,py)}=\text{pyridine}]$ spin-crossover solid 
\cite{Cofsa} which shows an abrupt thermally induced spin-crossover
transition centered at $126$K. For this system, the magnetic spin moments
corresponding to the HS and the LS states are $S_{\text{H}}=3/2$ and $S_{%
\text{L}}=1/2$, and the energy gap is 700 K \cite{Zarem}. In our
Hamiltonian, the parameter $h$ denotes $h=g_{\text{L}}\mu BS$, where $\mu $
is the Bohr magneton, and $B$ the applied magnetic field. Here $g_{\text{L}}$
is the Land\'{e} factor $g_{\text{L}}=2$ for both LS and HS states. For $%
\text{Co}^{II}(\text{H}_{2}\text{(fsa)}_{2}\text{en})\text{(py)}_{2}$, we
found that the shift $\Delta T_{\text{eq}}\approx -4.9$ K under a static
magnetic field of $32$ Tesla.

\subsubsection{$J\neq 0, h=0, K\neq 0, D\neq 0$ case.}

In this case, the HS units created thermally or optically 
interact through the exchange coupling $J$. We study effects of this
magnetic interaction on the charge transfer involving SC transition, under 
zero applied magnetic field. In MFT, the thermodynamical properties are
obtained by solving the following coupled self-consistent equations: 
\begin{equation}
m=q\tanh \beta zJm  \label{m4}
\end{equation}
and 
\begin{equation}
q=\frac{\cosh \beta zJm}{e^{-\beta \left( zKq-D+k_{\text{B}}T\ln 2g\right)
}+\cosh \beta zJm}.  \label{q3}
\end{equation}
Here, it is clear from the equation (\ref{m4}) that the system may order
magnetically, depending on the thermal behavior of the order
parameter $q$. Indeed, at Curie temperature $T_{\text{M}}$, we have 
\begin{equation}
q(T_{\text{M}})\beta _{M}zJ=1.  \label{qc}
\end{equation}

It is interesting to notice that the thermal behavior of this complicated
situation may be discussed through the comparison of the transition
temperatures of the magnetic and spin-crossover systems. At Curie
temperature $T_{\text{M}}$, the magnetization $m(T)$ shows a second-order
phase transition, and the HS fraction $q(T)$ exhibits a singularity.

Below $T_{\text{M}}$ spontaneous magnetization appears where the system is
globally in the HS phase. As the temperature decreases, the HS fraction
decreases smoothly and the magnetization disappears at $T_{\text{C}}(<T_{%
{\rm M}})$ with a second-order transition for small value of $J $ where $%
q(T) $ exhibits a singularity. When $J$ is large, the magnetization remains
until $T_{\text{eq}}$ and there disappears discontinuously at the first
order phase transition temperature.

\begin{figure}[h]
\caption{Temperature dependence of $m(T)$ (thin curves) and $q(T)$ (bold and
dashed curves) in the case $zK=90$K, $D=150$K and $g=10$ for the following
values of the magnetic interaction. From the bottom to the top: $zJ=48$K, $%
zJ=57.7$K, $zJ=58$K, $zJ=59$K, $zJ=61$K, and $zJ=63$K. The solid bold curve
denotes the HS fraction $q$ for $J=0$K, in which $m=0$. Note that the
magnetization $m(T)$ is equal to zero in the case $zJ=48K$.}
\label{fig-MFqm}
\end{figure}

When the SC and magnetic transition temperatures are of the same order of
magnitude, the magnetic 
system (i.e., $m(T)$) perturbs the ''lattice'' properties (i.e., q(T))
as seen in Fig. \ref{fig-MFqm}, where the bold solid line denotes the
temperature dependence of the order parameter $q$ for $J=0$. There all the
simulations have been performed varying the magnetic coupling $J$\ at fixed
values of $K$\ and $D$: $zK=90$\ K and $D=150$K. For small values of $J$, as 
$zJ=48$K, no magnetization appears and only smooth SC transition occurs.
When $zJ$ has values between $57.7$ and $59$ for example, we observe a
second-order re-entrant phase transition on the magnetization $m$ with the
transition temperatures $T_{\text{C}}$ and $T_{\text{M}}$. The associated HS
fraction shows singularity at each transition temperature. Increasing $zJ$,
such as $zJ=63$K, the magnetization presents a first-order transition at $T_{%
\text{C}}$ and a second-order transition at $T_{\text{M}}$. In such case,
the HS fraction exhibits a first-order transition at $T_{\text{C}}$ and a
singularity at $T_{\text{M}}$. Thus, it appears clearly that the magnetic
coupling causes the first-order transition in the elastic properties (i.e.,
in $q(T)$). In the following we discuss the three regions identified in Fig. 
\ref{fig-MFqm}:

(1) a high temperature region in which $q\simeq 1,$ $m=0$, corresponding to
a paramagnetic HS state;

(2) an intermediate region $T_{\text{C}}<T<T_{\text{M}}$, where $q>1/2,$ $%
m\neq 0$, indicating the existence of a predominant HS state with a
ferromagnetic order;

(3) at low temperature $T<T_{\text{C}}$, we have a phase of LS. There, we 
find two subregions. At relative high temperature region, $q$ is still
large, where many high spins remain, which are made of isolated magnetic
states $S=\pm 1$ in a sea of diamagnetic states $S=0$ where the system is
paramagnetic. On the other hand, at low temperature $q$ is very small and
the system is almost the perfect diamagnetic state. In this low temperature
phase in which $m=0$ and $q<1/2$, we obtain a reentrant magnetic phase.
Between the 
regions of (2) and (3), we find a first order SC phase transition driven by
the magnetic interaction. Due to the magnetic interaction, the high spin
fraction $q$ at the transition temperature $T_{C}$ is different from the
usual value $q=1/2$ obtained for $J=0$.

In the present situation, i.e. case (3), both $m$ and $q$ undergo phase
transitions. For the magnetization, when the temperature increases, magnetic
property changes as the dia- $\rightarrow $ para-$\rightarrow $
ferro-magnetic and then paramagnetic state. If $J$ is large, the low
temperature phase becomes ferromagnetic and then we have the change; the
ferro- $\rightarrow $ to para-magnetic states. Concerning the high spin{\bf %
\ }$q$, its temperature dependence can be either a first-order transition or
a continuous conversion. These results obtained by MFT will be studied also
by MC simulations in the next section.

To investigate the phase diagram of the SC system with magnetic
interactions; we first expand the equations (\ref{m4}) and (\ref{q3}) around
the magnetic transition temperature $T_{\text{C}}$(second-order case) or $T_{%
\text{M}}$ for which $m\approx 0.$ After some analysis, we arrive easily to
the following equation, giving the transition line in the ($zJ,T)$ plane

\begin{equation}
d-\frac{t_{c}}{j}=t_{c}\ln \left( 2g\frac{j-t_{c}}{t_{c}}\right) \text{ }
\label{phas_diag2}
\end{equation}
with $d=\frac{D}{zK},$ $t_{c}=\frac{T_{\text{C}}}{zK}$ and $j=\frac{J}{K}$.
Here the reduced critical temperatures corresponding to $T_{\text{C}}/zK$
and $T_{\text{M}}/zK$ are denoted by the same quantity $t_{c}$.

In order to observe the re-entrant phase transitions with the present model,
the function $j=f(t_{c},d)$ must have a minimum. Due to the self-consistent
structure of equation (\ref{phas_diag2}), it is difficult to predict
analytically the coordinates $\left( t_{\text{crit}},j_{\text{crit}}\right) $
of this minimum. Let us consider separately the left and right sides of the
equation (\ref{phas_diag2}), and denote these two quantities by $%
f_{1}(t_{c})=d-\frac{t_{c}}{j}$ and $f_{2}(t_{c})=t_{c}\ln \left( 2g\frac{%
j-t_{c}}{t_{c}}\right) $. At the critical temperature, we must have $%
f_{1}(t_{c})=f_{2}(t_{c})$ and $f_{1}^{^{\prime }}(t_{c})=f_{2}^{^{\prime
}}(t_{c}),$ where the $f^{^{\prime }}$ denotes the derivative of $f$ $\left(
t_{c}\right) .$ These two relations lead to the following transition line
for the critical points 
\begin{equation}
d+j_{\text{crit}}-1=j_{\text{crit}}\ln \left( 2g\frac{j_{\text{crit}}}{d}%
\right) .  \label{j_crit}
\end{equation}
Solving numerically the last equation, using $d=D/zK=150/90=5/3$ and $g=10$,
we obtain $j_{\text{crit}}\approx 0.64$, which gives $zJ_{\text{crit}}=57.6$
K. This value is in very good agreement with the thermal behavior of the
order parameters of Fig. \ref{fig-MFqm}. These results are summarized in the
phase diagram of a SC system including 
magnetic interaction between the HS units, depicted in Fig. \ref
{fig-MFT-reentrant}. A reentrant phase transition appears for $zJ_{\text{crit%
}}=57.6$K $<zJ<zJ_{0}=210$K where $j_{0}=2.33$. The magnetic phase appears
from $T_{\text{C}}$ to $T_{\text{M}}$. A magnetic second-order phase
transition occurs at $T_{\text{M}}$. On the other hand, a change from a
second-order to first-order magnetic phase transition at $T_{\text{C}}$
takes place while $J$ becomes large from $J_{{\rm crit}}$ to $J_{0}$. When
we choose larger $K$, $T_{{\rm eq}}$ becomes smaller and the transition at $%
T_{\text{C}}$ can become of first order (see next section).

\begin{figure}[h]
\caption{The phase diagram of a SC system with magnetic interactions in the (%
$\frac{J}{K}$, $\frac{T}{zK}$) plane. The coordinate $(j_{\text{crit}},t_{%
\text{crit}})$ denotes the end point of the reentrant phase transition. A
reentrant phase transition appears for $zJ_{\text{crit}}=57.6K<zJ<zJ_{0}=210$%
K and magnetic order appears from $T_{\text{C}}$ to $T_{\text M}$. A
magnetic second-order phase transition occurs at $T_{\text{M}}$. On the
other hand, magnetic transition at $T_{\text{M}}$ changes from second order
to first order with $J$. }
\label{fig-MFT-reentrant}
\end{figure}

Let us now discuss the case of small $J$ for which $m$ is always zero. In
this case, although the magnetic interaction may cause a metastable magnetic
phase at low temperatures, the system simply undergoes a transition from the
diamagnetic state (LS) to the paramagnetic state (HS). This situation
corresponds to the simple spin-crossover phenomena. The transition can be
smooth or discontinuous as the first order transition. The latter is the 
case of the experimental results observed in PBA, like Na$_{0.44}$Co[Fe(CN)$%
_{6}$]$_{0.73}$ . 2.7 H$_{2}$O \cite{Shimamoto}.

It is worth noting that many authors have investigated the phase diagram of
the original BEG model \cite{Ono}. Most of works on BEG model concerned the
case of antiferromagnetic biquadratic interaction ($K<0$) \cite
{Wang,Ono,Kaneyoshi} because it is expected a rich phase diagram due to
competing interactions. It was found that BEG model exhibits various
complicate phase transitions, such as successive phase transition,
re-entrant and double re-entrant double transition against the temperature.
Here, in the present study, both magnetic ($J$) and biquadratic ($K$)\
interactions are positive (ferro), and as long as these interactions are
''ferromagnetic'', the original BEG model does not show re-entrant phases.
However, in the present study we obtain a re-entrant behavior with $J>0$\
and $K>0$. On the other hand, the degeneracy of the HS state ($s=\pm 1$)\
combined to the ligand field $D$ act as an effective temperature dependent
''anisotropy'' $(D-k_{\text{B}}T\ln g)$\ which is strongly reduced when
temperature increases. As a consequence, in a rough approach, the original
phase diagram (in the plane $t_{{\bf c}}-j$ at constant $D$) of the original
BEG model (in which $g=1$) is changing depending on the temperature. This
effect constitutes the basic mechanism of a new competition leading to obtain
re-entrant phase transitions. 

\section{Monte Carlo Simulation}

In the previous section, we have studied the structure of the phase diagram
as a function of the lattice coupling $K$ and magnetic coupling $J$ in MFT.
In this section, we check the results obtained by MFT using a Monte Carlo
method where the effect of fluctuation is taken into account. Here, we adopt
an extended scheme of the Glauber dynamics because we study only static
properties of the model. We deal with three-dimensional system ($z=6$) with
20$\times $20$\times $20 sites through this section. The dynamical
properties such as the relaxation process after rapid changing of
temperature or photo-irradiation will be reported in the separate paper \cite
{next} where the types of dynamics will be carefully discussed.

Taking into account the degeneracy of the states, $u$ and $r$ for the low
and high spin states, respectively, we adopt the following transition
probability between the initial state $S$\ and the final state $S^{\prime }$%
: 
\begin{equation}
W(S\rightarrow S^{\prime })={\frac{P(S^{\prime })e^{-\beta E(S^{\prime })}}{%
P(0)e^{-\beta E(0)}+P(1)e^{-\beta E(1)}+P(-1)e^{-\beta E(-1)}}},
\label{Wrate}
\end{equation}
where $P(S)$ {\bf \ }is the degeneracy of the{\bf \ }electronic state,
represented by the fictitious spin $S$ ($S=\pm 1$ for the HS state and $S=0$
for the LS state). Thus, we have $P(0)=u$\ and $P(1)=P(-1)=r$. Here,  $E(S)$
is given by 
\begin{equation}
E(S)=-JS\sum_{j}S_{j}-hS-KS^{2}\sum_{j}S_{j}^{2}+\left( D-k_{%
\text{B}}T\ln g\right) S^{2}.  \label{E(S)}
\end{equation}
{\bf \ } This energy denotes the energy of the corresponding site, which can
be HS or LS and it includes elastic (trough the ''biquadratic'' interaction)
and magnetic contributions due to the interactions with the neighbors.  The
transition probability $W(S\rightarrow S^{\prime })$ satisfies the detailed
balance for the canonical distribution for the degenerate state.

we define the following quantity $m^{\prime }$ to study magnetic order in
the MC method 
\begin{equation}
m^{\prime }=\frac{\langle (\sum_{i}S_{i})^{2}\rangle }{N^{2}},
\label{mprime}
\end{equation}
where $\langle \cdot \cdot \cdot \rangle $ means MC average and $N$ is the
system size ($20^{3}$).

\begin{figure}[h]
\caption{(a) Temperature dependence of $q$ ($\bigcirc $) and $m^{\prime}$ ($%
\triangle $) obtained by the Monte Carlo simulation. (b) Temperature and
field dependence of $q$ for the first order phase transition. Circles ($%
\bigcirc $) denotes the case $h=0$, ($\triangle $) $h=9$ and ($\Box$) $h=18$%
. (c) Temperature and field dependence of $m^{\prime}$ for the first order
phase transition. Note that the magnetization shows non-monotonic behavior
for nonzero $h$. }
\label{fig-qT}
\end{figure}
In Figs. \ref{fig-qT}, we show the temperature dependence of $q$ ($\bigcirc $%
) and $m^{\prime}$ ($\triangle $). Here the temperature is raised up from $%
T=5$ to $80$K sequentially, and is reduced to $T=5$K again. In Fig.\ref
{fig-qT}(a), smooth change of $q$ and no change of $m^{\prime}$ is depicted
for $J=0$, $h=0$, $g=10$, $K=15$K and $D=150$K. The transition temperature ($%
T_{{\rm eq}}=35.1{\rm K}$) is larger than the critical temperature ($T_{%
\text{I}}=16.9 {\rm K}$) of the corresponding Ising model and no phase
transition occurs. We find that the plotted circles for both upward and
downward processes are well superimposed at each temperature. In Fig.\ref
{fig-qT} (b) and (c), we plot $q$ and $m^{\prime}$ in the case of the first
order phase transition, respectively, where $g=10$, $K=28$K, and $D=150$K,
giving $T_{{\rm eq}}=22.0{\rm K}$ ($<T_{\text{I}}=31.5$ {\rm K}), for $J=0$, 
$h=0$ ($\bigcirc $) and for $h\neq 0$ ( $\triangle$, $\Box$ ) .

We performed relatively short steps (200MCS) at each temperature for Figs 6,
which is enough to get equilibrium for case (a), leading well visible
hysteresis loops of $q$ for case (b).

The shape of the hysteresis changes with the duration time in principle.
Indeed, if we change the temperature infinitesimally slow, the system would
change in a quasi-static way and the hysteresis would disappear. However, in
practice, the nature of metastability is robust and the shape of the
hysteresis changes little in 
the case with $5000$ MCS at each temperature. In the Fig.\ref{fig-qT}(b) we
also confirm the field-dependence of the shape of the hysteresis obtained in
the previous section by MFT. Triangles ($\triangle$) denote the case $h=9$
and squares ($\Box$) $h=18$. Shift of hysteresis curve to the low
temperature side is observed and the shift is more prominent in downward
process.

\begin{figure}[h]
\caption{Temperature dependence of $q$($\bigcirc$) and $m^{\prime}$($%
\triangle$) for $K=15$K, $D=150$K, $g=10$, and $J=12$K obtained by the Monte
Carlo simulation. Inset shows the heat capacity ($\bullet$). Note that the
magnetic interaction drives the first-order SC transition.}
\label{fig-qmT}
\end{figure}

Here, we study the effect of magnetic interaction on the spin-crossover
phase transition. We consider the case where the magnetic interaction $J$ is
nonzero. If we include weak magnetic interaction, the qualitative nature of
the order parameter $q$ does not change, although the interaction causes a
metastable branch of magnetically ordered state at low temperatures. This
effect will be reported when we study the dynamical properties of the system 
\cite{next}. Here, we investigate the change of equilibrium phases which
were studied in the previous 
section in MFT. In order to change the structure of phase diagram we need a
strong magnetic interaction which is comparable to $K $. In Figs.\ref
{fig-qmT}, temperature dependence of $q$ and $m^{\prime }$ is depicted for $%
K=15$(K) and $D=150$(K), where $T_{{\rm eq}}>T_{\text{I}}$, with $J=12$(K).
At each temperature, $2000$ MCS are performed to obtain the data. We find
that the magnetization appears at intermediate temperatures. Decreasing the
temperature, magnetization appears continuously at the critical temperature $%
T_{{\rm M}}\simeq 43$K where the phase transition is of the second order as
was predicted in MFT, and vanishes discontinuously in the first order phase
transition at $T_{\text{C}}\simeq 32$K, where the HS fraction changes
discontinuously.

\begin{figure}[h]
\caption{Phase diagrams in the coordinate $(J/K,T/zK)$ obtained by the Monte
Carlo simulation. Closed squares denote $T_{\text M}$, closed triangles $T_{%
\text C}$, and closed circles the SC transitions. (a) for $K=15$K, $D=150$K,
and $g=10$ (b) for $K=22$K, $D=150$K, and $g=10$.}
\label{fig-TCM}
\end{figure}

Now, we study dependence of the critical temperatures $T_{M}$ and $T_{\text{C%
}}$ as a function of $J$ for the case $K=15$(K), $D=150$(K) and $g=10$,
where $T_{{\rm eq}}>T_{{\rm I}}$. In Fig.\ref{fig-TCM}(a), the dependence is
depicted, where the closed circles denote the temperature at which $q=1/2$.
The transitions depicted by the closed circles occur smoothly. The closed
triangles denote $T_{\text{C}}$ and the close boxes $T_{M}$. The magnetic
transition at $T_{\text{C}}$ changes from second order to first order with $%
J $ between $J/K=0.73$ and $J/K=0.8$.

As another example, we also give the phase diagram in Fig.\ref{fig-TCM}(b)
for the case $K=22$(K) and $D=150$(K), where $T_{{\rm eq} }<T_{{\rm I}}$.
Then the closed circles denote the phase boundary of the first order phase
transition. Technically we obtain $T_{{\rm eq}}$ and $T_{\text{C}}$ of the
first order phase transition as follows. We prepare an initial condition
where half of the system is in LS state ($S=0$) and the other half is in the
magnetic HS state ($S=1$). We perform MC simulations with different random
number sequences for given $J$ and observe the relaxation of $q$ and %
$m^{\prime }$. When all samples go to LS (HS para, HS magnetic) 
phase, we regard the parameter set as belonging to the LS (HS para, HS
magnetic) phase. We could determine the border within the precision of the
size of the closed circles and closed triangles. These phase diagrams are
very similar to those obtained by MFT in the previous section. It is
interesting to study the intermediate parameter region between (a) and (b),
where the line connecting the closed circles with the smooth transition
(case (a)) changes to that with the first order phase transition (case (b)).

\section{Summary and discussion}

The phase diagram of the system with the spin-crossover phase transition and
the magnetic phase transition has been studied using an extended BEG model
with degenerated levels by means of the mean-field theory and Monte Carlo
simulations. This model can be a model for both SC and PBA materials in the
framework of a common description. The present model can be also seen as an
extension to 3 states of the Wajnflasz and Pick model for spin-transition 
\cite{Pick}. This unified approach contains both ''elastic'' and magnetic
interactions between the molecular species. These two kind of 
interactions obey to different symmetries, and then involve two different
order parameters. Indeed, the terms responsible for the SC phenomena  are
the temperature-dependent ligand field $\left( D-k_{\text{B}}T\ln g\right)
S_{i}^{2}$ and the biquadratic $-KS_{i}^{2}S_{j}^{2}$ interaction, which
modulate the energy between HS and LS 
states, leading to the gradual or first-order transition. On the other hand,
the magnetic interaction $-JS_{i}S_{j}$, assumed here as ferromagnetic, acts
only between the HS units, which breaks the symmetry, and leads to the
second-order or first-order magnetic transition when it competes with the SC
transition.

We found that when the magnetic interaction is weak, it does not cause a
magnetic phase transition at even very low temperature at equilibrium. It
only causes a metastable magnetic state at very low temperature. This is
consistent with the observation in PBA. The properties of the metastable
state will be reported in the study of dynamics \cite{next}. 
It has also been found that the magnetic interaction causes a magnetic order
in the HS phase at intermediate temperatures, i.e. $T_{\text{C}}<T<T_{{\rm M}%
}$. The magnetic order disappears at a lower temperature than $T_{\text{C}}$.

Even in the condition of $K$ and $D$ that $q(T)$ changes smoothly when $J=0$%
, i.e., $T_{{\rm eq}}>T_{{\rm I}}$, the magnetic interaction drives the spin
crossover transition which causes the first-order transition for large $J$.

As we have referred in the text, the dynamical properties of the present
model are also very interesting and important in the light of recent
experimental extension of photo magnetization.\cite
{Hash1,Hash2,Photo1,Photo2} They would be related to a description of the
photo-magnetic behavior of PBA, in particular, of the relaxation properties
of the photo-induced magnetic state. There, the relaxation of two coupled
magnetic and elastic order parameters, which have different time scales, is
involved.\cite{next}

\label{sec:matome} \acknowledgements
The present work was supported by Grant-in-Aid for Scientific Research from
Ministry of Education, Science, Sports, Culture, and Technology of Japan,
and Centre National de la Recherche Scientifique (CNRS, PICS\ Japon N${%
{}^\circ}2272$) of France. One of the authors (K. Boukheddaden) is grateful
for financial support for the collaboration expenses from the Grants. The
numerical calculation is supported by the supercomputer center of the
institute of the solid state physics of Tokyo university, which is also
deeply acknowledged.


\begin{references}
\bibitem[*]{concurrent}  Also at International Center For Young Scientists,
National Institute for Materials Science, Tsukuba, Ibaraki 305-0044, Japan.





\bibitem{Ammeter}  J.H.\ Ammeter, Nouv. J. Chimie 4, 631 (1980)

\bibitem{Kahn}  O. Kahn, Molecular Magnetism, VCH, New York, 1993.

\bibitem{Varret}  F.\ Varret, M.\ Nogu\`{e}s, A.\ Goujon, in: J.\ Miller,
M.\ Drillon (Eds.), Magnetism: Molecules to Materials, vol.\ 2, Wiley-VCH,
2001, p. 257.

\bibitem{Sorai}  M.\ Sorai, Chem.\ Soc.\ Jpn, 74, 2223 (2001).

\bibitem{Gutlich}  P.\ G\"{u}tlich, Y.\ Garcia, T.\ Wo\"{i}ke, Coord.\
Chem.\ Rev. 219, 839\ (2001).

\bibitem{varret2}  \ F.\ Varret, A.\ Bleuzen, K.\ Boukheddaden, A.\
Bousseksou, E.\ Codjovi, C.\ Enachescu, A.\ Goujon, J.\ Linares, N.\
Menendez, M.\ Verdaguer, Proc. Pure Appl. Chem.\ Rev. (2002) in press.

\bibitem{Verdaguer}  M. Verdaguer, Science 272, 698 (1996).

\bibitem{Sato1}  H.W. Liu, K. Matsuda, Z.Z. Gu, K. Takahashi, A.L. Cui, R.
Nakajima, A. Fujishima, O. Sato, Phys. Rev. Lett. 90, 167403 (2003).

\bibitem{Freysz}  E. Freysz, S. Montant, S. L\'{e}tatrd, J.F. L\'{e}tard,
Chem. Phys. Lett. 394, 318 (2003).

\bibitem{tof}  H.\ Toflund, Coord.\ Chem. Rev. 94, 67 (1989).

\bibitem{Koenig}  E.\ K\"{o}nig, Struct. Bonding 76, 51 (1991).

\bibitem{gut2}  P. G{\"{u}}tlich, A. Hauser, H.\ Spiering, Angew. Chem. Int.
Ed. Engl. 33, 2024 (1994).

\bibitem{Sato}  O.\ Sato, T.\ Iyoda, A.\ Fujishima, K.\ Hashimoto, Science
272, 704 (1996).

\bibitem{Bleuzen}  A.\ Bleuzen, C.\ Lomenech, V.\ Escax, F.\ Villain, F.\
Varret, C.\ Cartier dit Moulin, M.\ Verdaguer, J.\ Amer.\ Chem.\ Soc. 122,
6648 (2000).

\bibitem{Goujon}  A.\ Goujon, O.\ Roubeau, M.\ Nogu\`{e}s, F.\ Varret, A.\
Dolbecq, M.\ Verdaguer, Eur.\ Phys.\ J.\ B. 14, 1145 (2000).

\bibitem{Kojima}  N. Kojima, W. Aoki, M. Itoi, Y. Ono, M. Sato, Y. Kobayashi
and Yu. Maeda, Sokid State Commum. 120 165 (2001).

\bibitem{Miya-Kojima}  S. Miyashita and N. Kojima, Prog. Thore. Phys. 109
729 (2003).

\bibitem{Kambara}  T.\ Kambara, J.\ Chem.\ Phys. 70, 4199 (1979).

\bibitem{Jeftic}  J.\ Jeftic, H.\ Romstedt, A.\ Hauser, J.\ Phys.\ Chem.\
Solids 57, 1743 (1996).

\bibitem{Boillot}  M.-L. Boillot, J.\ Zarembovitch, J.-P. Ities, A.\ Polian,
E.\ Bourdet, J.G.\ Haasnoot, New.\ J.\ Chem. 25, 313 (2002).

\bibitem{Jeftic2}  J.\ Jeftic, N.\ Menendez, A.\ Wack, E.\ Codjovi, J.\
Linares, A.\ Goujon, G.\ Hamel, S.\ Klotz, G.\ Syfosse, F.\ Varret, Meas.\
Sci. Technol. 10, 1059 (1999).

\bibitem{AncaSava}  A. Sava, C. Enachescu, A. Stancu, K. Boukheddaden, E.
Codjovi, I. Maurin, F. Varret, Journal of Optoelectronics and Advanced
Materials 5, 977 (2003).

\bibitem{Kojima2}  Y. Kobayashi, M. Itoi, N. Kojima and K. Asai, J. Phys.
Soc. jpn. 71 3016 (2002).

\bibitem{Pick}  J. Wajnflasz, Phys. Status Solidi 40 537 (1970). J.
Wajnflasz and R. Pick, J. Phys. Colloq. France 32, C1 (1971).

\bibitem{Doniach}  S. Doniach, J. Chem. Phys. 68, 11 (1978). M. Nielsen, L.
Miao, J.H. Ipsen, O.G. Mouritsen and M.J. Zuckermann, Phys. Rev. E. 54, 6889
(1996).

\bibitem{kbo} K. Boukheddaden, I. Shteto, B. H\^{o}o, F.\ Varret, 
Phys. Rev. B {\bf 62}, 14796 (2000); ibid. 14806 (2000).

\bibitem{Spiering1}  H. Spiering, E. Meissner, H. K\"{o}ppen, E.W. M\"{u}%
ller, P. G\"{u}tlich, Chem. Phys. 68, 65 (1982). N. Willenbacher, H.
Spiering, J. Phys. C. 21, 1423 (1988). H. Spiering, N. Willenbacher, J.
Phys. Cond. Matter. 1, 10089 (1989).

\bibitem{Jamil1}  J. Nasser, K. Boukheddaden, J. Linares, Eur. Phys. J. B.
39, 219 (2004).

\bibitem{Jamil}  J.\ Nasser, Eur.\ Phys.\ J.\ B. 21, 3 (2001).

\bibitem{nishino3}  M.\ Nishino, S.\ Miyashita, K.\ Boukheddaden, J.\ Chem.\
Phys. 118, 10 (2003).

\bibitem{Az1}  A. Bousseksou, J. Nasser, J. Linares, K. Boukheddaden, and F.
Varret, J. Phys. I France 2, 1381 (1992).

\bibitem{koppen}  H.\ K\"{o}ppen, E.W.\ M\"{u}ller, C.P.\ K\"{o}hler, H.\
Spiering, E.\ Meissner, P.\ G\"{u}tlich, Chem. Phys.\ lett.\ 91, 348 (1982).

\bibitem{nishino4}  M. Nishino, K. Boukheddaden, S. Miyashita, and F.
Varret, Phys. Rev. B 68, 224402 (2003).

\bibitem{nishino1}  M. Nishino, K. Yamaguchi and S. Miyashita, Phys. Rev. B
58,  9303 (1998).

\bibitem{nishino2}  M. Nishino and S. Miyashita, Phys. Rev. B 63,  174404
(2001).

\bibitem{BC1}  M. Blume, Phys. Rev. 141, 517 (1966).

\bibitem{BC2}  H. W. Capel, Physica (Amsterdam) 32, 966 (1966); 33, 295
(1967); 37, 423 (1967).

\bibitem{next}  M.\ Nishino, K.\ Boukheddaden, S.\ Miyashita, F.\ Varret, in
preparation.

\bibitem{BEG}  M.\ Blume, V.J.\ Emery, R.B.\ Griffiths, Phys.\ Rev.\ A.\ 4,
1071 (1971). K.\ Kasono, I.\ Ono Z.\ Phys.\ B.-Condensed Matter 88, 205
(1992).

\bibitem{PTP_Kamel}  K. Boukheddaden, Prog. Theor. Phys. 112, 205 (2004).

\bibitem{PRB}  I. Shteto, K. Boukheddaden, F. Varret, Phys. Rev. E.60, 5139
(1999).

\bibitem{Sivardiere}  J. Sivardi\`{e}re and J.\ Lazerowitch Phys.\ Rev.\ A.\
11, 2079 (1975); J. Sivardi\`{e}re and J.\ Lazerowitch Phys.\ Rev.\ A.\ 11,
2090 (1975); J.\ Sivardi\`{e}re and J\ Lazerowitch, Phys.\ Rev.\ A. 11, 2101
(1975).

\bibitem{Luty}  T. Luty, in Relaxations of Excited States and Photo-Induced
Phase Transitions, edited by K. Nasu (Springer-Verlag, Berlin, 1997), p. 142.

\bibitem{Koshihara}  M. H. Lem\'{e}e-Cailleau, M. Le Cointe, H. Cailleau, T.
Luty, F. Moussa, J. Roos, D. Brinkmann, B. Toudic, C. Ayache, and N. Karl,
Phys.\ Rev.\ Lett. 79, 1690 (1997)

\bibitem{Koshihara2}  S. Koshihara, Y. Tokura, T. Mitani, G. Saito, and T.
Koda, Phys. Rev. B 42, 6853 (1990); S. Koshihara et al., Synth. Met. 70,
1225 (1995). 

\bibitem{Varret3}  F.\ Varret, A.\ Goujon, K.\ Boukheddaden et al., Mol.\
Cryst.\ Liq.\ Cryst.\ 379, 333 (2002).

\bibitem{Galina}  G.S.\ Matouzenko, G.\ Molnar, N.\ Br\'{e}fuel, et al.,
Chem.\ Mater.\ 15, 550 (2003).

\bibitem{LSV}  J. Linares, H. Spiering, F. Varret, Eur. J. Phys. B 10, 271
(1999).

\bibitem{1Dkbo}  K. Boukheddaden, J. Linares, F. Varret, Eur. J. Phys. B.
15, 317 (2000). 

\bibitem{Mukamel}  D. Mukamel, M.\ Blume, Phys.\ Rev.\ B.\ 10, 610 (1974)

\bibitem{Koshi1}  Y. Ogawa, T. Ishikawa, S. Koshihara, K. Boukheddaden, and
F. Varret, Phys. Rev. B. 66, 073104 (2002)

\bibitem{Azpulse}  A.\ Bousseksou, N.\ Negre, M.\ Goiran, L.\ Salmon, J.-P.\
Tuchagues, M.-L.\ Boillot, K.\ Boukheddaden, F.\ Varret, Eur.\ Phys.\ J.\ B.
13, 451 (2000).

\bibitem{PRBpulse}  A. Bousseksou, K. Boukheddaden, M. Goiran, C. Consejo,
M-L. Boillot, and J-P. Tuchagues, Phys. Rev. B 65, 172412 (2002).

\bibitem{Cofsa}  N.\ Tohira, H.\ Okawa, S.\ Kida, Chem.\ Lett.\ 1979, 683
(1979).

\bibitem{Zarem}  \ J.\ Zarembovitch, New.\ J.\ Chem. 16, 225 (1992).

\bibitem{Shimamoto}  N.\ Shimamoto, S.\ Ohkoshi, O.\ Sato, K.\ Hashimoto,
Inorg.\ Chem. 41, 678 (2002).

\bibitem{Ono}  I.\ Ono, J.\ Phys.\ Soc.\ Jpn.\ C8, 1541\ (1988).

\bibitem{Wang}  Y.L.\ Wang, C.\ Wentworth, J.\ Appl.\ Phys. 621, 4411\
(1987).

\bibitem{Kaneyoshi}  T.\ Kaneyoshi, J.\ Phys.\ Soc.\ Jpn. 56, 4199 (1987).

\bibitem{Hash1}  S.\ Ohkoshi, K.\ Hashimoto, J.\ Photochem. Photobiol., C\
2, 71(2001).

\bibitem{Hash2}  T.\ Yokoyama, K.\ Okamoto, T.\ Ohta, S.\ Ohkoshi, K.\
Hashimoto, Phys.\ Rev.\ B. 65, 64438 (2002).

\bibitem{Photo1}  H.\ Tokoro, S.\ Ohkoshi, K.\ Hashimoto, Appl. Phys.\
Lett.\ 82, 1245 (2003).

\bibitem{Photo2}  V. Escax, A. Bleuzen, J. P. Iti\'{e}, P. Munsch, F.
Varret, and M. Verdaguer, J. Phys. Chem. B 107, 4763 (2003).





























\end{references}
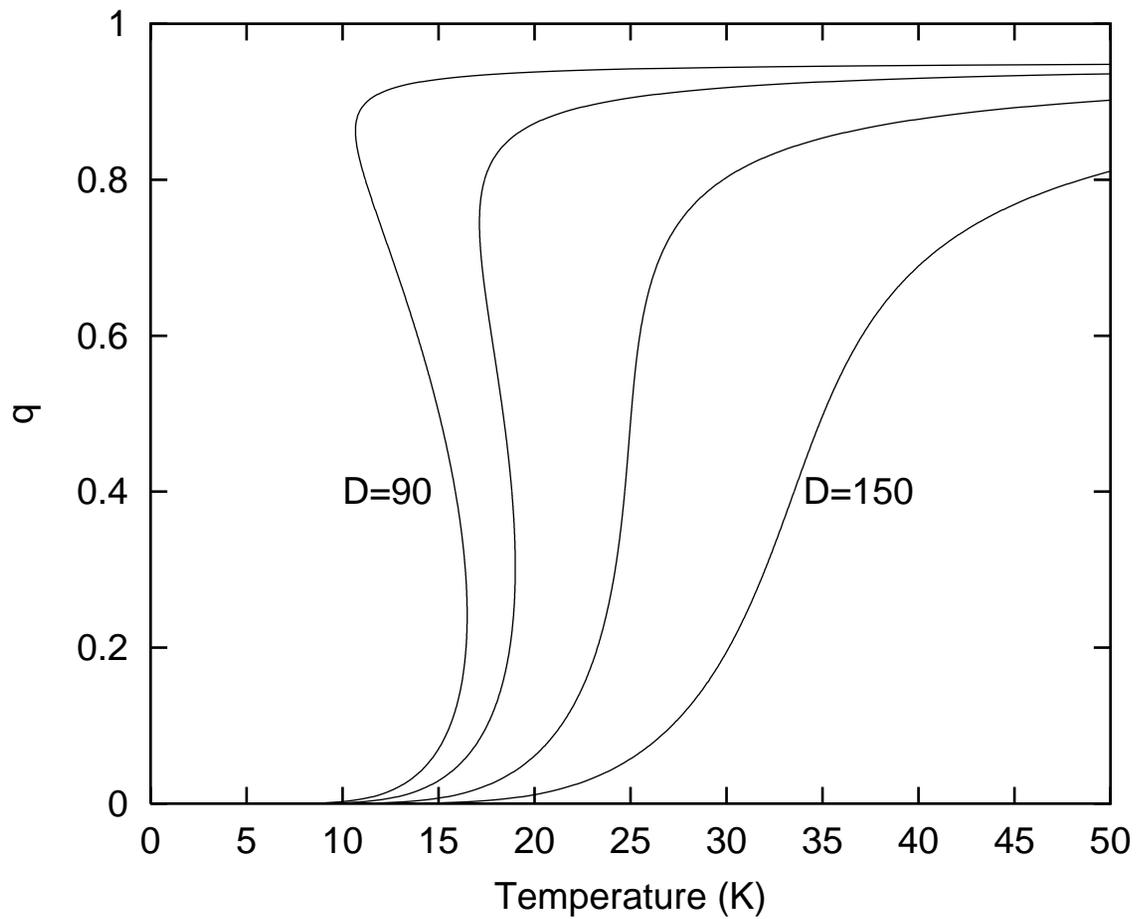
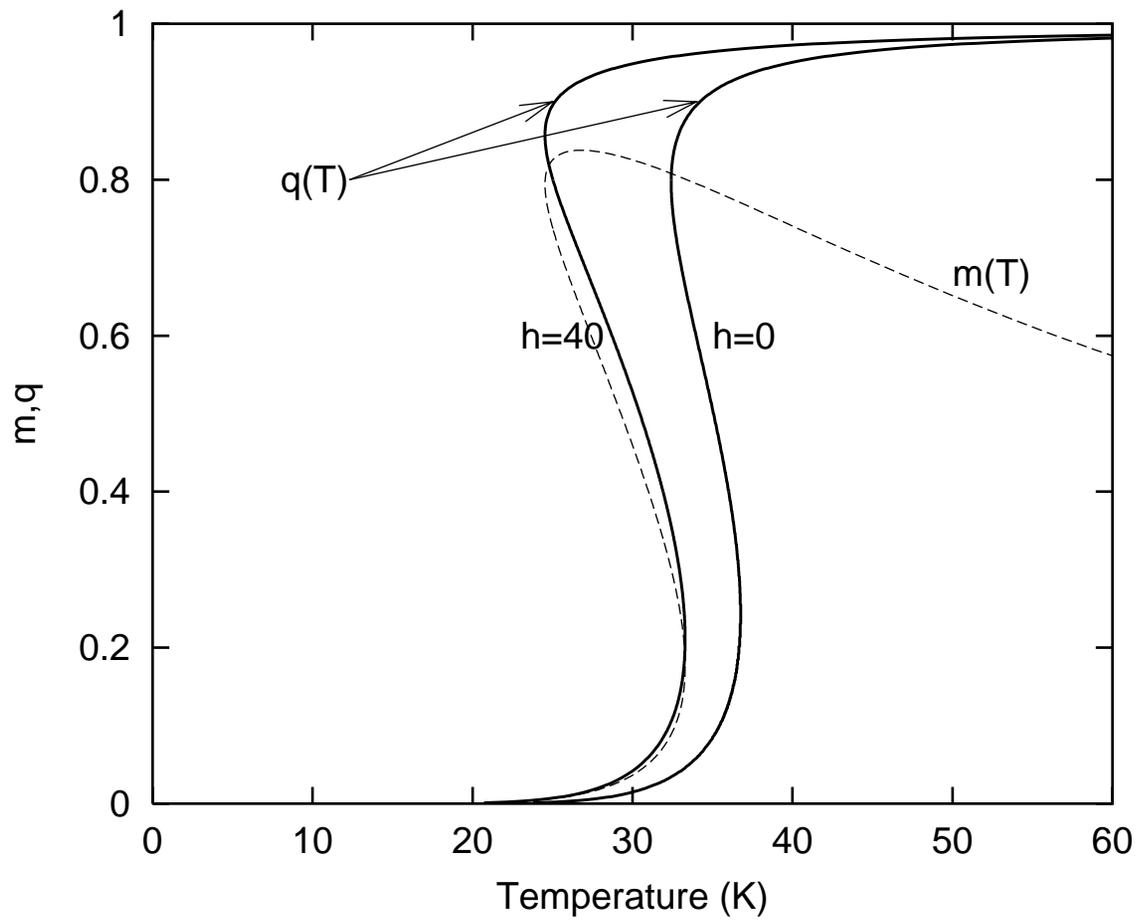
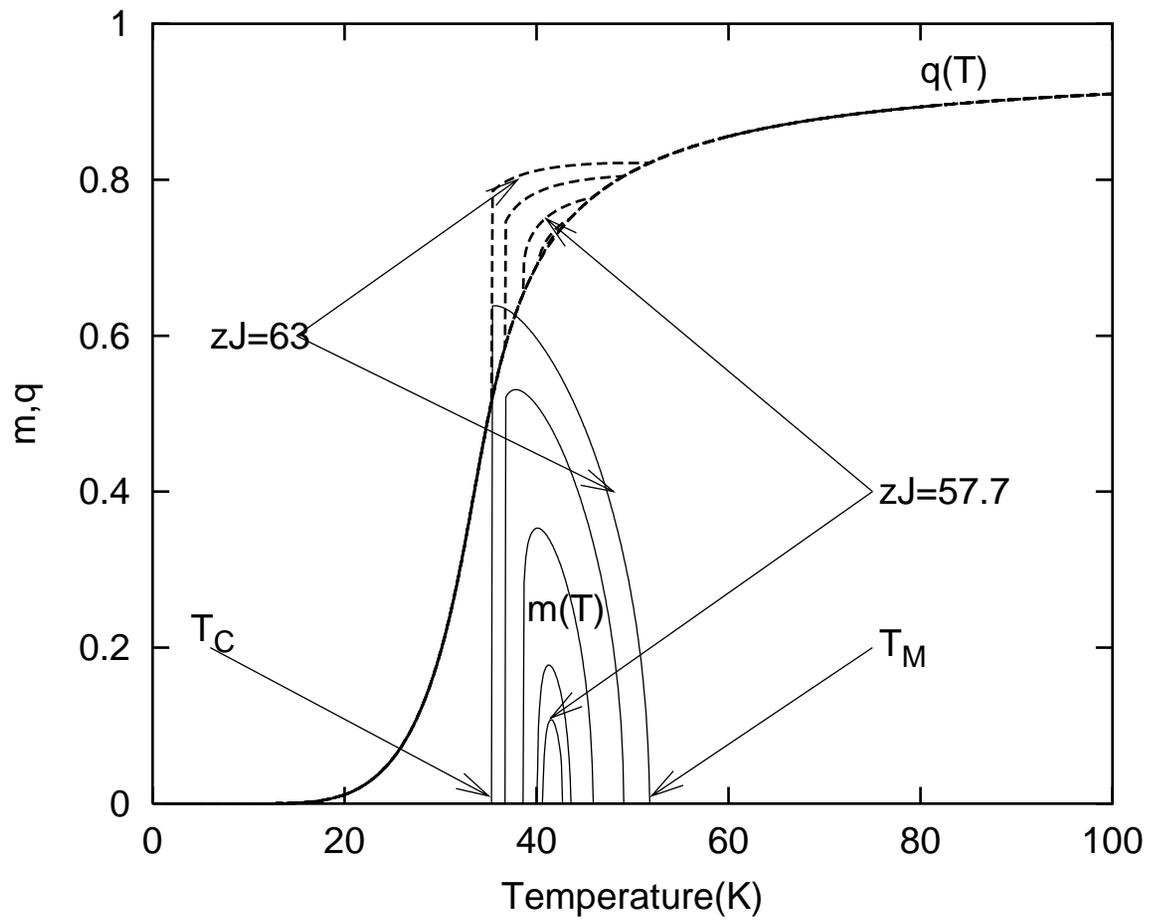
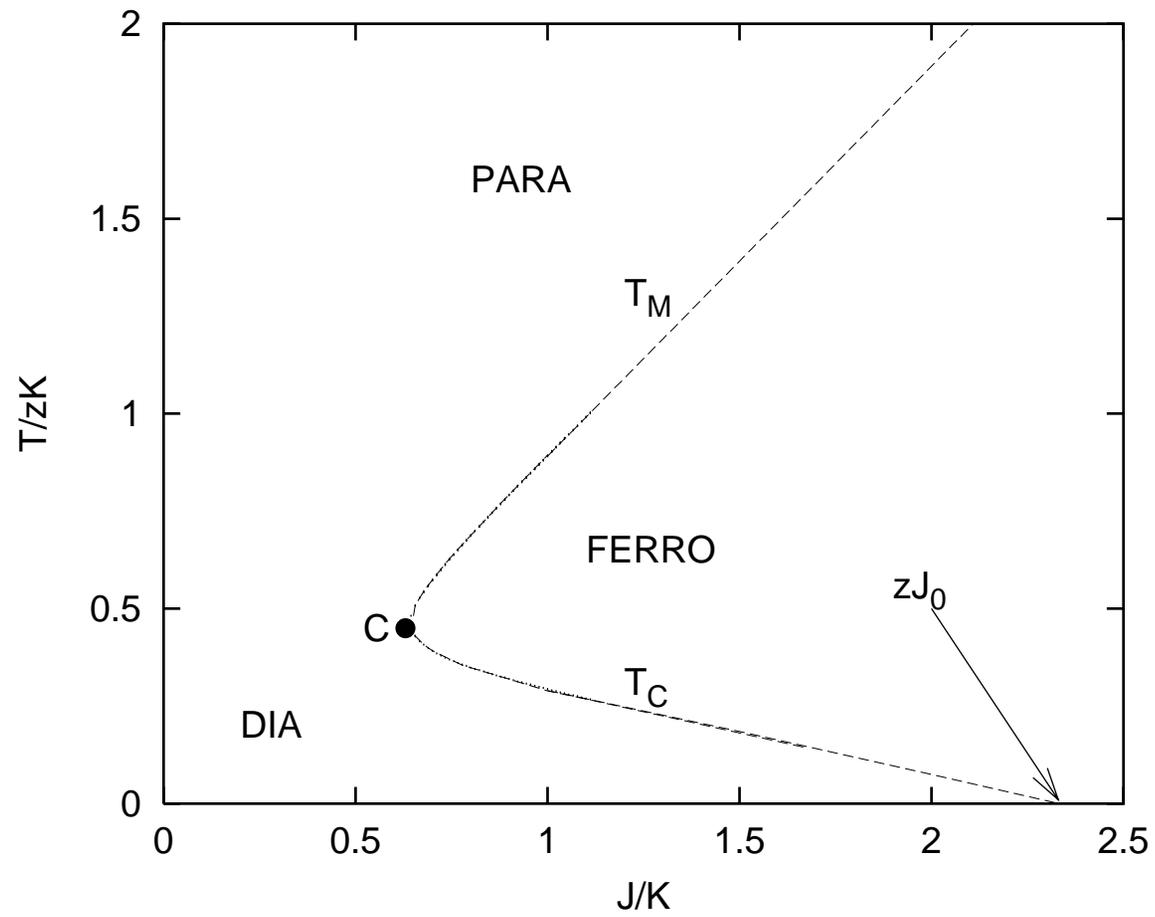
\end{document}